# You Watch, You Give, and You Engage:
# A Study of Live Streaming Practices in China

Zhicong Lu, Haijun Xia, Seongkook Heo, and Daniel Wigdor
University of Toronto, Ontario, Canada
{luzhc | haijunxia | seongkook | daniel}@dgp.toronto.edu

## ABSTRACT
Despite gaining traction in North America, live streaming has not reached the popularity it has in China, where live-streaming has a tremendous impact on the social behaviors of users. To better understand this socio-technological phenomenon, we conducted a mixed methods study of live streaming practices in China. We present the results of an online survey of 527 live streaming users, focusing on their broadcasting or viewing practices and the experiences they find most engaging. We also interviewed 14 active users to explore their motivations and experiences. Our data revealed the different categories of content that was broadcasted and how varying aspects of this content engaged viewers. We also gained insight into the role reward systems and fan group-chat play in engaging users, while also finding evidence that both viewers and streamers desire deeper channels and mechanisms for interaction in addition to the commenting, gifting, and fan groups that are available today.

### Author Keywords
Live streaming; social media; user engagement; social network.

### ACM Classification Keywords
H.5.m. Information interfaces and presentation (e.g., HCI): Miscellaneous.

## INTRODUCTION
In recent years, the proliferation of mobile devices equipped with high definition cameras and high-speed internet has led to a surge of individuals making live streams. These *live streams* are supported by software that enables anyone to share their experiences at live events, hold weekly talk shows, and more. Twitch.tv, the popular gameplay-based live streaming platform has more than 2.2 million active streamers per month [8,35]. Despite the popularity of Twitch.tv, live streaming in North America has only recently begun to enter into mainstream culture, through the advent of Facebook Live, YouTube Live, and Periscope [7,18,33]. Although a few research projects have focused on understanding the live streaming phenomenon, they have been largely confined to understanding North American usage [7,8,18,26,33]. The present work, however, casts a lens on a country and culture that has almost ubiquitous live streaming usage: China.

The continuous growth of live streaming in China has resulted in more than 200 million viewers watching streamers perform live each night on more than 200 live streaming platforms, creating an estimated 5-billion-dollar industry in 2017 [23]. Chinese live streams differ greatly in content, style, and form compared to those in North America and Europe. Previous studies of US and Canadian live streaming have found that live streaming was almost exclusively for live events [7,33] or sharing among close friends [18]. In contrast, Chinese users utilize live streaming for a wide array of uses, such as *pan-entertainment* (i.e. so-called "showroom performances" of singing, dancing, music instruments, and talk shows hosted and performed by individual streamers), e-commerce, personal knowledge sharing, and personal experience sharing [42]. Although differences in live streaming activities have been identified, there is little understanding as to why these differences exist, what makes live streaming activities so engaging and popular in China, and what we as the designers of live streaming platforms can learn from their use.

To better understand this socio-technological phenomenon, we conducted a mixed methods exploration that included an online survey (N = 527) and interviews (N = 14) with regular users in China. The survey queried the motivations behind engaging with live streams and live streamers, in addition to the 'rewards' viewers provide to streamers, and other methods of interaction they have with streamers off the live streaming platforms. Our results revealed that Chinese viewers are more interested and engaged in live streams about the personal experiences of strangers than they were in the experiences of their friends, live events or civic content, contrasted prior results on North American live streaming usage. We also found that reward-based systems and fan groups on instant messaging apps further afford social interactions between viewers and streamers and facilitate community building. We provide an overview of the opportunities and challenges of live streaming exposed by this research, to inform the design of future platforms and services that support social live streaming.

## BACKGROUND AND RELATED WORK

We first provide background of live streaming in China, then present related work regarding vloggers on YouTube, video game live streaming, and general live streaming.

### Live Streaming in China

In China, live streaming began when the Internet users started repurposing the public video chat room services of YY [44] for performative live streaming [2]. Most live streaming platforms gradually began to provide mobile versions of their live streaming services. As of April 2017, there have been more than 200 live streaming platforms in China [42], with 324 million live streaming users [5]. The top 9 popular platforms all have over 2 million daily active users [11]. Due to the popularity of live streaming, 'streamer' has become a new profession in China. Corporations such as Tencent provide facility support and training for full time streamers who have the potential to become popular [37]. These professional streamers go live almost every day and live streaming acts as a primary source of income for them.

### YouTube and Vloggers

YouTube allows for the sharing of user-generated video content, and provides a road to fame for many vloggers who use YouTube to share their life and experiences. Wesch [38] analyzed the self-awareness generated by vlogging, and provided evidence of context collapse [20] on YouTube. The commenting system on YouTube was analyzed by Siersdorfer [30] and it was found that most comments were concentrated on a small portion of the most popular videos. Some research in HCI has also explored how to craft a conversational experience and promote social interactions through YouTube [9]. Although relevant, the present work thus seeks to understand how real-time, synchronized interaction during large-scale live streaming influences viewer engagement and the social interactions between viewers and streamers.

### Video Game Live Streaming (on Twitch.tv)

Hamilton et al. [8] found that live streaming communities form around shared identities from streams' content and viewers' shared experiences. Pellicone et al. [21] conducted a qualitative study of an online forum about game streaming and found that a key attribute of streaming is the development of a unique attitude and persona as a gamer. Sjoblom et al. [32] investigated why viewers watch others play video games, and showed that tension release, social integration, affective motivations, and information seeking impacted the hours spent watching, subscription behavior, and the number of streamers watched. Similarly, Gross et al. [6] investigated how motivations influence user's gratifications on watching live streaming on Twitch.tv, finding that time spent influences the money spent on the platform. Sjoblom et al. [31] also studied how video game genres and content types influenced viewer gratification in live streaming, and showed that archetypal structure of the stream is more important than genre of games being streamed. Although relevant, the focus on gaming as a shared experience makes this sort of live streaming a unique community. We focused our research efforts on the entire set of live streaming uses.

### General Live Streaming

Live streaming has attracted increased attention in the HCI community due to the popularity of mobile live streaming applications. Juhlin et al. [12] explored early generations of mobile live streaming apps Bambuser and Qik, finding that users could not find appropriate topics to stream and streams suffered from poor image quality. Dougherty [4] explored the use of Qik, showing that 11% of videos had civic importance, such as journalistic and activist value, e.g., activist meetings. Tang et al. [33] conducted a mixed methods study of early users of live streaming on Meerkat and Periscope, finding that the broadcasted live streams consisted a diverse range of activities, and the motivations of the streamers were mostly for branding purposes.

Other work explored the demographics of users on live streaming services and what makes such stream engaging. Scheibe et al. [28] analyzed information production and reception behavior on YouNow, revealing that YouNow streams were mostly made by adolescents for adolescents. Lottridge et al. [18] studied the live streaming practices and motivations of teens, highlighting that teens were streaming like taking a long-form selfie, showing one's self and life, to engage with small group of friends. Work by Hu et al. [10] did attempt to understand why users kept watching on live streaming platforms, and found that self-identification with the broadcaster and viewing group increased the motivation to keep watching. Haimson et al. [7] studied what makes live events engaging by comparing Facebook Live and Snapchat Live Stories. They found that immersion, immediacy, interaction, and sociality were important to the engagement of watching live events.

This existing research explored the different contexts where live streaming is used, streamers' behavior, and user engagement in North America and Europe, but little research has explored the practices of live streaming in China, where there are more daily active users [5], and more professional and regular streamers [11]. Work by Zhou [42] attempted to understand live streaming behaviors in China, revealing information about users' demographics, usage, and perceptions of live streaming, however, he did not explore what content was being viewed, how viewers interact with streamers, and what factors contributed to viewers' engagement. To better inform the design of future live streaming interfaces, we seek to dig deeper into the phenomenon of lives streaming in China and understand how the practices, engagement, and interactions in North America differ from those in China.

## METHODS

Inspired by the prevalence and popularity of live streaming in China, we explored the following research questions:

**RQ1**: What are the practices and motivations for watching and conducting live streams in China? How do they differ from those in North America?

**RQ2**: What type of interactions between viewers and streamers take place? How are these interactions different from those in North America?

**RQ3**: What mechanisms exist to reward streamers or show one's appreciation for their content? How do such mechanisms affect engagement with a streamer?

**RQ4**: What are the factors contributing to a viewer's engagement [24] while watching live streams? How do they differ from previous work?

### Data Collection: Survey

To explore these questions, we adapted the methods of Rader et al. [27] and Baumer et al. [1], who used an online survey-based methodology to collect stories about issues related to computer security and Facebook non-use, respectively. We developed an online questionnaire in a similar manner, using two types of questions. The questionnaire was first developed in English with the research team, then translated into Chinese by one of the native Chinese authors and validated by another.

The first line of questioning probed the live streaming experience of viewers. These questions included mostly yes/no, multiple choice, or 5 point Likert-style questions and focused on understanding whether the respondent is a viewer, a streamer, or both, how long s/he watches a live stream, which features of live streaming platforms s/he uses most often, what factors of streamers attract him/her to keep on watching them, and other similar questions. The items were adapted from existing survey study on live streaming in China [42], as well as previous research on user engagement [7] (e.g., entertainment, sociality, information and interactivity). English and Chinese versions of the questionnaire, along with complete anonymized response data, can be found in an appendix to this submission.

The second set of questions were open-ended and probed live streaming experiences. Similar to Rader et al. [27] and Baumer et al. [1], all respondents were asked to tell a story about a live stream they watched in the past three weeks that they thought was the most engaging, creative, or enjoyable. They were also encouraged to provide a link or a screenshot of the stream they mentioned. We also asked them to briefly describe interactions or features they wish were provided in current live streaming platforms.

*Recruitment*

To avoid our results being biased by just one platform or certain groups of users, we used services from SoJump.com to recruit respondents, which is a firm specializing in recruiting study participants in China. The survey was active for 2 weeks in July 2017. We received 902 completed surveys, which we reviewed using manual inspection and SoJump's data cleaning tools. In total, 375 responses were removed from the data set for one or more of the following reasons: not being a live streaming user (257), failing our "trap" questions (12), time of completion (55), patterned responses (36), or for providing fake or gibberish text responses (15); this cleaning method followed [7]. This left 527 completed responses for analysis. Each respondent who finished the questionnaire received a cash payment. Our analysis of results was conducted in Chinese, and translated to enable broader collaboration with the team and reporting in this paper.

### Data Collection: Interviews

To understand live streaming practices and some findings from the survey data more deeply, we interviewed 14 participants who participated in the survey and agreed to participate in a follow-up interview (Table 1). The interviews were conducted remotely using video calls during August 2017. Each interview lasted about 50 minutes and participants were provided with a 50 CNY honorarium for their time. The interviews were semi-structured, with questions about participants' current experiences regarding live streaming applications. Interviews were conducted in Mandarin, audio-taped, and transcribed by the author who conducted the interviews.

| ID | Gender | Age | Years of Use | Has Streamed Themselves | Frequency of Use in times/week | Location |
|---|---|---|---|---|---|---|
| P1 | F | 21 | 3 | Yes | 7+ | Southwest |
| P2 | M | 29 | 0.5 | No | 3 | Midwest |
| P3 | M | 34 | 1 | Yes | 3 | Southwest |
| P4 | F | 27 | 2 | Yes | 2 | Mideast |
| P5 | M | 33 | 0.5 | No | 4 | Midwest |
| P6 | M | 28 | 1 | No | 3 | Northwest |
| P7 | M | 39 | 2 | No | 2 | Mideast |
| P8 | F | 31 | 1.5 | No | 7+ | Southeast |
| P9 | M | 35 | 0.5+ | Yes | 3 | Northeast |
| P10 | M | 23 | 1 | No | 5 | Southeast |
| P11 | M | 26 | 1.5 | Yes | 4 | Mideast |
| P12 | M | 32 | 1 | No | 3 | Northwest |
| P13 | F | 22 | 2 | Yes | 7+ | Southwest |
| P14 | F | 24 | 1 | No | 4 | Northeast |

Table 1. Summary of interview participants' demographic information and location (within a 9-sector grid of China).

### Analysis

Responses to the open-ended questions were analyzed using an open coding method [3]. The two native Chinese authors coded the first 10% of responses and met to gain consensus on their codes. One author coded the remaining responses and met with the second coder to reach agreement. All the codes were then translated into English and were discussed by the broader research team using affinity diagramming to group and find emerging themes. The transcription of the interviews was analyzed in the same way as the open-ended questions in the survey.

We employed factor analysis to cluster sets of questions about engagement, in order to find significant differences in the patterns of the data. Factor loadings were derived through principal component decomposition [15] and then rotated using the varimax (orthogonal) rotation, following a similar approach as in [13,18].

**RESULTS**

The results have been organized based on the themes that emerged from the affinity diagramming exercise. We first present a demographic profile of the respondents and then detail users' motivations and general practices while using live streaming apps, what the reward system afforded, and the importance of fan groups on instant messaging applications, the characteristics engaging streams had in common, and the factors contributing to engagement. We present a fusion of the open-ended questions in the survey and the interviews to explain the findings in greater detail.

**Profile of Respondents**

The questionnaire took on average 20 minutes to complete (M=20.30, SD = 15.49), and was completed by 243 males and 282 females (Figure 1). Respondents reported being full-time students, managers, developers, researchers, teachers, professionals, and so on. The age and education level distributions align with previous market report [2] and research [42] about live streaming users in China.

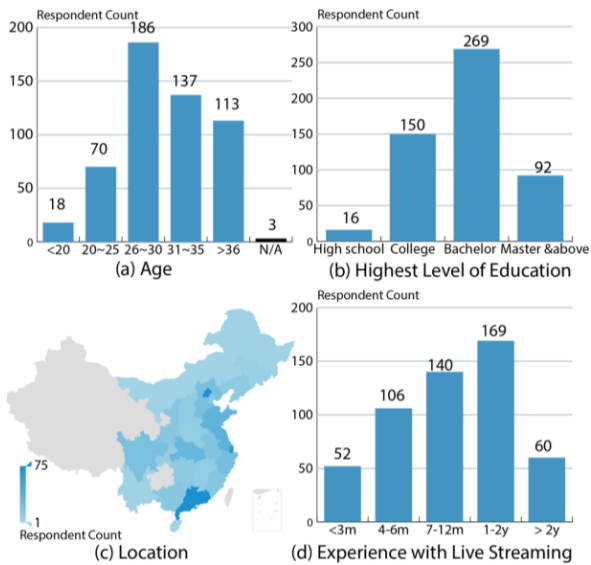

**Figure 1. Distribution of age, highest level of education, location, and experience with live streaming.**

Of all the respondents, 363 only watched others' streams but did not stream themselves, 14 streamed themselves but seldom watched others, 45 spend more time streaming themselves than watching others, and 105 spend more time watching others than streaming themselves.

Respondents reported using multiple live streaming platforms, with an average of 4 platforms per respondent (M=3.9, SD=3.1; Table 1). We note that similar to previous reports on live streaming in China [2,42], the most popular apps were all well represented in our respondents responses, and most respondents (93%) used mobile versions of the live streaming services. Although most platforms have a variety of genres of streams, viewers chose to use different platforms for different content, for example, "*Douyu is more for games and performance while Kuai Shou is for dramatic performance*" (P7).

| App Name | # of Users | % | App Name | # of Users | % |
|---|---|---|---|---|---|
| Douyu.tv | 287 | 54.46% | Panda.tv | 131 | 24.86% |
| YY Live | 263 | 49.91% | Yi Zhi Bo | 104 | 19.73% |
| Momo | 210 | 39.85% | Long Zhu | 88 | 16.70% |
| Inke | 200 | 37.95% | Qi Xiu (x.pps.tv) | 81 | 15.37% |
| Kuai Shou | 198 | 37.57% | Lai Feng | 74 | 14.04% |
| Hua Jiao | 188 | 35.67% | Ha Ni (ihani.tv) | 49 | 9.30% |
| Hu Ya | 157 | 29.79% | Xian Dan | 44 | 8.35% |

**Table 2. Respondents' usage of live streaming services.**

Respondents reported that they found live streams of interest via content shared by friends (60%), via browsing featured streams (59%), or via the leaderboard of streamers (57%). This indicates that views could be skewed to the most popular fraction of streamers who were featured on the platform or had lots of fans, similar to findings of [33].

The average time spent watching a single live stream was 62.0 minutes (M=62.0, SD=33.6), compared to the average video length on YouTube (4 minutes and 20 seconds) [22]. Respondents also spent on average 7.2 hours (M=7.2, SD=5.0) per week watching live streams, demonstrating that most respondents were regular live streaming users.

**Motivations for Watching**

More than half (69%) of respondents reported that they watch live streaming to relax, while 55% reported being attracted by certain streamer. Killing time, making more friends, communicating with others and sharing life, sharing their point of view, finding a community, or desiring to gain new knowledge were also reported as important motivations for watching live streams.

The comments from interviewees also provided other motivations. For example, some reported watching live streaming because they felt lonely and wanted to seek comfort and the feeling of company. Others sought out experiences that were different from their own social life, "*We seldom have parties in China. Sometimes I really want to escape from my close-tie or the circle of my fellows, and live streaming provides a perfect place for that*" (P1). Some also mentioned motivations for seeking specific information, "*Before I went to Hong Kong for a business trip, I intentionally watched a stream from a local streamer for some advice and her own recommendations for food*" (P14).

**Activities and Topics**

With regards to the content that was most favored by respondents, there were a variety of activities that viewers had a high interest in watching (Figure 2). We now elaborate our understanding by further analyzing the stories from the survey, and mainly focus on topics and activities not explored in previous literature.

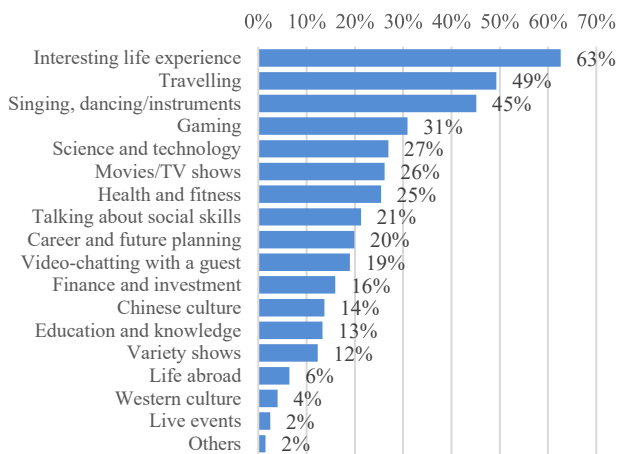

**Figure 2. Responses to "What kinds of live streams do you enjoy watching the most (Choose up to 5)".**

*Sharing Personal Experiences*
Unlike many North American streams, streams that were conversational in nature, whereby the conversation was between a streamer and a group of strangers, were very popular with respondents (131 responses in total). Sixty-eight of the responses were about streams where the streamer was a non-celebrity and shared their real-life story or personal experiences. In other streams, conversations centered around performing or playing video games, but the emphasis was on engaging in the conversation with others who were not one's own friends. This desire to talk with others who are not part of one's traditional social circle differs from previous live streaming research [18].

Most streams that were told about focused on meaningful, serious topics, such as dealing with the pressure from work or family, work-life balance, career planning and development, dealing with relationships, marriage, raising the next generation, or supporting one's parents. Given the pressure that most Chinese adults from 26 to 35 years old face from their career and their family's expectations for marriage and descendants, live streaming seems to provide a medium through which such individuals can have a social support system without disclosing too much information or involving those close to them. "*It's really beneficial to have interactions with streamer and viewers. I can release the pressure from my daily life, my work, my family, children, parents, and high pace of life of modern society, through discussion with the streamer and viewers during the stream. Peers around me cannot fully understand my pain*" (S354).

With streams that were more serious in nature, some streamers announced the topic to be covered during the stream in advance. As noted by S163, "*The streamer notified all the fans in advance that the topic would be marriage, which attracted a lot of fans. The streamer told about experiences of her friends, with unique and thorough perspectives. Feeling her emotion and the realness of the story, I was inspired by the stream greatly.*" This gave viewers more time to think about the topic and potentially inspire deeper conversations – much unlike the live and unpredictable nature of most North American streams.

There were some conversational streams mentioned that were focused on streaming itself, i.e., the streamer shared his/her own life experience, how s/he became a popular streamer, and maintained and got more fans. Viewers appreciated the inspiration and positive attitudes conveyed by such chatting (S343), as it gave them opportunities to know more about streamer because they did not appear to be into streaming to gain popularity or fame, but enjoyed the activity itself and were willing to share this joy and their process with others. In many cases, this stimulated one's willingness to follow the streamer, as mentioned by P13.

*Performance*
Watching amateur or semi-professional spontaneous performances was also a very popular streaming activity to watch (i.e., 125 responses in total, with 85 "singing" responses, 19 "talk show" responses, 13 "dancing" responses, and others included magic tricks, playing music instruments, and performing unique skills like *xiangsheng* (i.e., Chinese comedic crosstalk) and ventriloquism).

Most streams in this category took place indoors, and were performed by beautiful women or men who were amateurs. They had other professions, even high education level, but really liked performing for fun, to attract fans, or to "*realizing early dreams in life*" (S121). Respondents reported appreciating the personality, taste, temper, and the humor of streamers, and enjoyed chatting with the streamer on topics about his/her personal live or current affairs. "*Although she is not a professional singer, her performance is of high quality and makes me happy and relieved*" (P4).

Viewers also appreciated the live and unedited nature of these performances, and enjoyed the interactions with the performer and other viewers, e.g., asking the streamer to sing a specific song and getting a response in real time, or listening to the stories related to the songs and commenting on them. The improvisational nature of such streams has resulted in a new form of improvisational music, *Han Mai* (a form of rapping) [17]. "*He once rewarded some of his fans by writing lyrics of a Han Mai with stories of his fans and performing it, which moved a lot of his fans*" (S127).

*Knowledge Sharing*
Another popular set of topics for live streams were those relating to acquiring new knowledge (92 responses, among which "Cooking" had 33 responses). Different from Ask Me Anything style streams [33], where viewers ask questions about certain topic and the streamer answers them, these streams were more like lectures or a training sessions, where the streamer talked about the topic, showing the procedure or essential information using slides or other media, and answered questions from viewers. Most streamers who do this also sought to provide a series weekly of streams on certain topics, which made the streams more like a series of live, online courses.

Such topics often included, formal knowledge (e.g., language learning, graduate school entrance exam preparation, college-level mathematics, psychology, finance and investment, Chinese traditional culture or techniques, etc.) or informal knowledge (e.g., cooking, interview skills, pick up skills, antique evaluation, skin-care, personal health, baby-care, fitness, DIY etc.). Cooking was especially popular, due to the oft complexity of Chinese culinary skills, as noted by P6, "*Since some Chinese recipes don't provide accurate instruction on amount of ingredients to be used and accurate time of the cooking, it's really helpful to watch the streamer cooking in real time, and even cook along with the streamer.*"

The streamers usually share such knowledge in a relaxing and entertaining style rather than preaching. The viewers appreciated this kind of streaming, because they could learn something in a fun, interactive way (P3, P10). Viewers also appreciated the possibility for customizing practices of certain knowledge enabled by live streaming. As noted by our respondents, "*There're different methods for keeping health using Chinese medicine, but it's hard to choose the appropriate methods and right medicine for me*" (S501). Live streaming enabled viewers to learn from the experience of the streamer and other viewers, and to ask customized questions and get them answered in real time.

### Videogaming
Similar to North American live streaming on services like Twitch.tv, watching others play video games was another common, slightly less popular activity in China (i.e., 88 responses). Most games were the same as on Twitch.tv (e.g., League of Legend, The Sims), while some were very popular Chinese mobile games (e.g., Honor of Kings).

The motivations for watching video-game play were similar to North America: to learn game play skills and strategies, to meet other gamers, and to enjoy the game without investing too much time playing and mastering it. These motivations echoed previous research about the importance of forming communities among game play [8] and cultivating the streamer's own game play attitude [26].

The characteristics of such streams that made them engaging related to the impressive skills and strategies of the gamer, the humorous way they interacted with viewers, and their temperament and positive attitude. One respondent recalled this story about the values of his favorite streamer. "*[The streamer] once started a reward for fans, asking them to send emails to him about their wishes and he'll fulfill some of it. Many fans sent emails, but one of them impressed me a lot. He wanted to be a volunteer teacher at the charity school which the streamer has donated to, and asked him for recommendation. Although that school was in less-developed region and doesn't provide decent pay, he would like to be a teacher there.*" (S29). This story shows the strong tie that developed between the streamer and his fans, and the potential greater social impact that streamers often have on their audience.

Unlike a number of North American streams, many respondents reported that game streamers sometimes also stream other activities, such as singing performances or outdoor activities, thereby raising their engagement with the stream. "*[The streamer] regularly changes her content since it would be boring for some fans if she always streams the same game. It's impressive that she sings really well and has a very good taste in music.*" (P10) This echoed the findings of live streaming among teens [18], but differs from those about video game live streaming.

### Outdoor and Travelling
A new category that emerged in the responses was the desire to see streamers travel and be outdoors (i.e., 67 responses in total, with 26 of them about outdoor activities, such as surviving in the wild, adventures, hunting, hiking, or climbing mountains). Viewers liked the thrilling feeling and the novelty of such streams, since such streams took place in locations where viewers seldom go to, and they knew it was real and unedited because it was live. They reported holding their breath for the streamer and worrying about their safety, while trying to learn some outdoor skills, as noted by S404, "*I watched a stream about downhill climbing at a 300-metre-high cliff, which was thrilling and exciting. The streamer told us about how to do downhill climbing and how to warm up through his detailed demonstration. I learnt a lot from it*".

Outdoor streams also provide lens to view society from the streamer's perspective, while without live streaming such behavior may be socially awkward. Some streams involve streamers streaming outdoors while interacting with passers-by, who were in their natural settings and reacted authentically. Viewers liked being able to observe these people and discuss the interactions that resulted. "*It makes me realize how people from other places differ from me in behaviors, attitudes, and values, which helps me to understand our society more thoroughly*" (P11).

### E-commerce
Another novel category of streaming that was popular came from live streams that were selling (mainly on Taobao, the Chinese largest e-commerce website). Although akin to North American infomercials, this is a unique category which is seldom seen in previous literature. Products ranging from books, clothes, health-care products, skin-care products, to even cars were demonstrated and shown in live streams. When asked if they would like to try the goods recommended by the streamer, respondents reported an average likelihood of 3.64 (M=3.64, SD = 0.77), which indicated that it was likely that they would try. This suggests that some viewers were influenced by the streams and may trust the streamer's recommendations.

Unlike the traditional pre-taped infomercial, respondents appreciated the ability to view the product directly, ask the streamer to show different perspectives of the product, see the process of the crafting of the goods, and ask detailed questions about the product. P4, who is a frequent consumer on Taobao, also noted that live streaming "*gives me chance to get to know other consumers in the stream, and I prefer to trust them rather than the streamer, as long as they don't seem to be shill*". Such e-commerce streaming highlights that live streaming is beginning to integrate with other mobile apps viewers use frequently, thus bringing additional opportunities and challenges for these services. It also suggests a potential avenue for the next generation of infomercials in North America.

*Miscellaneous*

There were also other stream content types mentioned, such as "Exaggerating behaviors", "Pets", "Celebrity", and "Competitions". One "Exaggerating behaviors" that emerged was about a streamer who live streamed themselves gorging on food, "*At first, I was curious about how she could eat so much. As she ate more and more, I became nervous and scared, since I knew it was live and it looked so real. After she finished I would get a sense of satisfaction, feeling that eating was the happiest thing in the world.*" (P1). It was the surprise, nervous feelings, and satisfaction that the viewers incurred that kept them interested in this exaggerating behavior.

**Interactions Between Viewers and Streamers**

In terms of the interactive features that respondents use, commenting to express self-emotion, to communicate with the streamer or other viewers, or to send free gifts to the streamer, were frequently reported (Figure 3).

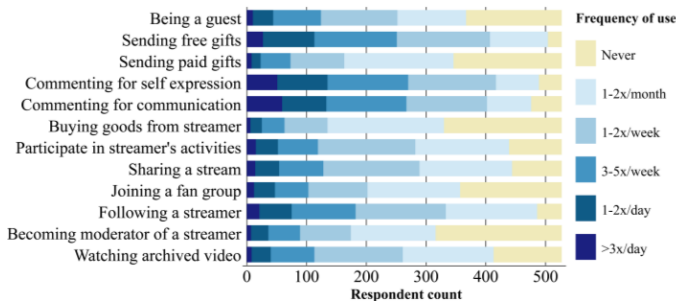

**Figure 3. Respondents' reported use of live streaming features in the month preceding the survey.**

Besides commenting, becoming a guest and video-chatting with a streamer during the stream was an important way for viewers to interact with the streamer. Although it is not used frequently, some interviewees reported that they liked the direct and lasting communication channel and the ability for collaboration during the stream that it afforded. Others found the guest sharing his/her own stories and performing together with the streamer "*fun and fresh*" (S91). Being able to communicate directly can facilitate deeper conversations, as S513 noted, "*during video-chatting with her I realized that we had some friends in common but we haven't met in person. After that we became very good friends in real life*". Participants did express some privacy concerns about being a guest, "*someone can identify me even if I just reveal my voice*" (P10).

Although we did not focus our study on streamers, 164 respondents regularly broadcasted, and 6 of our interview participants had streamed, with 2 of them being regular streamers. When asked about motivations for broadcasting themselves, most reported that they wanted to share their life with others, show their talents, or make more friends. The need for branding, which is shown to be the main motivation of streamers on Periscope and Meerkat [33], was a small part of their motivation. Most reported seeking to satisfy their social needs, "*I stream mostly to share my experience with others who share the same interest with me, and to get a sense of accompany when being alone*" (P14).

*Gifting*

On Chinese live streaming platforms, sending free gifts are similar to "liking" a social media post [29] or "hearting" a stream [33], but with different visual designs and more opportunities to show one's personality. Live streaming platforms in China also offer the opportunity for further interaction between a viewer and a streamer, whereby a viewer can purchase and send a virtual gift to a streamer during the live stream. When a viewer sends a virtual gift, a message is displayed to all viewers of the stream, thus serving as a public broadcast of one's admiration and appreciation for the streamer. As noted by P14 "*the floating and flashing gifts on the screen were just like fluorescent sticks of a concert, which echoed the streamer's performance*". To encourage repeated giving, a leaderboard of the top gift senders is visible to all viewers, and all viewers are notified when a top gift sender joins the stream. Streamers can receive a portion of the value of the gifts, and the rest goes to the platform.

Surprisingly, many respondents reported sending paid gifts (66%). This was commonly done because the viewers liked the content, they wanted to show support for the streamer, they wanted to reward the streamer, or they simply liked the streamer. Gifting with fancy animations provided viewers with a means of self-expression. P13 noted that "*sometimes sending virtual gifts works as my special way of saying hi to the streamer*". P3 noted, "*Sometimes I send gifts just as a creative Danmu to show my agreeance with the streamer or other viewers*". This usage is similar to the use of emoji and stickers [43], which express rich and subtle emotions that cannot easily be conveyed by text alone.

Paid virtual gifts also allow streamers to wade through the large number of messages that may be sent to a streamer more easily, because viewers who want to demand attention can pay for virtual gifts. Some viewers sent paid gifts with the intention of catching the streamer's attention to lower the chance of their comments being ignored by the streamer, especially when asking specific questions or sending a special request to the streamer. P3 noted that "*Since I paid for it, it would stay much longer than text comments. It's really important if you want to catch the attention of the streamer when there are over 10,000 concurrent viewers sending comments*". In this way, they no longer passively watched the stream, but were active participants, influencing the content of the stream more so than those who do not send gifts. For example, P13 noted that "*I once sent a gift to let the streamer sing a song I like best of her*". Viewers who sent gifts can also develop a further connection with the streamer, such as becoming the moderator of the chatroom and getting a chance to have private contact with the streamer. One story told that "*I really liked her performance and paid for a lot small gifts for her. Then I became the 'guardian' of her chatroom and we became good friends afterwards*" (S152).

Our data also revealed several problems with gifting, which may result in viewers not being willing to pay for virtual gifts. One problem is that the threshold needed to satisfy a streamer can be raised if s/he continually receives expensive gifts. Such behavior, however, is visible to all viewers, which makes some unwilling to pay for more gifts. As noted by P2, "*Now I don't send gifts very often. When someone had sent gifts worth of 1000 RMB, I think it doesn't make sense to send gifts worth of 50 RMB to the streamer. She won't pay extra attention anyway.*" Another problem is that the benefits of sending gifts do not accumulate effectively, as noted by P13, "*when I have sent several gifts to the streamer and developed friendship with him, sending more gifts doesn't have more benefits to me*". Some streamers even treat gift-senders differently from non-gift-senders, which most of our participants were aware of and said would impact their impression of a streamer.

In addition to providing gifts within a streaming platform, 51% of respondents indicated that they had rewarded streamers through other external channels, such as WeChat Red Packets [39], money transfer through AliPay (similar to PayPal), or by physically mailing the streamer gifts. P14 noted, "*I have ordered food online late at night for my favorite streamer who was working hard on her streams*". This behavior demonstrates that as viewers and streamers develop a friendship, they begin to disclose more personal information, such as online accounts or addresses.

Gifting is also not a one-way exchange. Several respondents revealed that sometimes streamers reward fans during the stream. As noted by S193, "*the streamer was wearing a costume with a lot of red packets on it. We could get a chance to pick one packet if we correctly answered a question about content in her previous streams. A lot of fans were participating*". Two participants (P3, P13) mentioned that they had rewarded their fans in the group chat by sending money through red packets in WeChat or QQ [39]. Interestingly, they did not find that group members were reluctant to open red packets, even though previous research has suggested that people are protective while getting involved in financial activities with strangers [39]. Some streamers also send gifts to each other during a stream, as a way of social exchange and maintaining "*guanxi*", the ties between individuals fostered through exchanges of favors [40], on the platform.

*Fan Groups*
Unlike North American live streams, where interaction is typically restricted only to the platform or Twitter, streamers also interact with viewers through other channels. Instead of using the built-in community in live streaming apps, or leveraging social networks like Weibo, most streamers build fan groups on instant messaging apps such as WeChat or QQ, which is an integral part of daily life in China [21,43]. The fan groups are group chats, with features to support multiple modalities of messaging beyond text, including emoji, stickers, voice, image, video messages or URLs to share other web pages. Such functionality is currently missing in the commenting systems of live streaming apps. Streamers usually put the information of fan groups in the banner of their profile pages or show it directly on the screen during their streams, so viewers can join the group by searching for the group chat ID or scanning the group QR code.

One reason a streamer (P3) mentioned for not choosing the built-in group chat function is that "*There are a lot of spams or even trolls in the group. On the other hand, I use WeChat and QQ more privately, and I check messages in WeChat or QQ more frequently than live streaming apps*".

As mentioned by the two other participants who regularly stream, fan groups were used to send notifications about going live, asking for feedback about certain streams, or asking the viewers about their interest. The fan groups are thought to be more suitable for these purposes, as noted by P7, "*I normally don't ask for feedback during my stream, since there are too many people sending comments, but few would give constructive suggestions. People in my group chat, on the other hand, care about me much more, and I really appreciate the suggestions they gave me*".

Over half of the respondents (56.4%) had joined the fan groups of streamers. The most common reasons why they joined were to communicate with streamers and other viewers, to gain more information about the streamer, or simply to make more friends in the group. As P14 noted, "*I made a friend in the fan group of my favorite streamer. We regularly chatted in the group and went along well, so one day I sent a request to him to add him as a contact in WeChat and we became friends afterwards*".

**Elements of an Engaging Stream**
Based on our analysis of our data, several elements were found to increase one's engagement with a live stream.

*Streamer Personality and Skills*
Many positive characteristics of streamers repeatedly emerged in our data, including welcoming, good-tempered, polite, honest, authentic, kind, empathetic, patient, grateful, impartial, hard-working and positive. Viewers also liked good voice, humor, story-telling skills, inter-personal skills, skills to handle awkwardness, and skills to deal with trolls wisely. Viewer's identification of these characteristics requires a deeper level of attention to the stream.

*Atmosphere*
Many responses mentioned that the most engaging streams have a good atmosphere that may come from including music of good taste, classy background decorations, light of the setting, the relaxing and pressure-free style of the streamer, and the polite behaviors and friendly comments of other viewers. A good atmosphere may help viewers stay watching longer and be more engaged in the stream.

*Novelty*
Viewers also seem to be more engaged in things they have not experienced, places they have not been to, knowledge they do not have access to before, and unexpected results. Therefore, they like to watch streams about traveling or outdoor activities, or streams with exaggerated behaviors.

*Authenticity*
Viewers also like the authentic nature of live streaming. Although some planning of the stream is essential and beneficial, viewers do not like the content and settings to be too artificial, but prefer them to be "close to life". As mentioned by P5, who liked watching the live behind-the-scenes streams of a popular variety show on TV, live streaming "*provides a chance for me to know something that would be cut on the final version of the show on TV*".

**Factor Analysis**
Data from 14 questions of 5-point Likert scale related to live streaming engagement was used to help find latent factors that explain the variance found in the questionnaire responses and the importance of these factors to the engagement of viewers. Kaiser-Meyer Olkin measure of sampling adequacy is above 0.6 ($KMO = .844$), and Bartlett's test of sphericity was significant ($\chi^2 (91) = 1926.9, p < .05$), which indicated that the sample was factorable. The factor analysis yielded 4 factors with eigenvalues more than 1 (Table 3). Each item was included within the factor with which it had the highest factor loading, and we averaged items in each factor.

"Content and Form of the Stream" (M=4.29, SD=0.53) was found to be the most important factor for engagement in a stream. Thus, the diverse content streamed on Chinese live streaming platforms are the key to their popularity. Different people can be attracted to streamers with different talents and personalities, diverse content, novel forms of performance, or previously unknown knowledge.

"Aesthetics" (M=3.84, SD=0.63) indicates the importance of visual elements, such as the UI, the appearance of the streamers, and even decorations of the background. This is also related to "Atmosphere" in the Engagements section.

"Communication with Others" (M=3.66, SD = 0.70) is also thought to be important, highlighting the importance of social interactions with others during live streams (through commenting) and beyond (through fan groups).

"Emotional Reactions" (M=3.53, SD=0.68) is an interesting factor, since its composition indicates that gift sending can be quite emotional. Several stories in our data echoed this, as S382 noted, "*At first I thought he was close to losing the game (LoL), but suddenly he came back and won the game by his adept operation. I was so impressed and spent nearly the expense of food for half a month to send him gifts*".

| Factors | Items (all in the form of "X is very important to the engagement of the stream") | Factor loading |
|---|---|---|
| Content and Form of the Stream Alpha = .64 (inter-item correlation) | Talents and personality of the streamer | 0.661 |
| | Interesting or meaningful content | 0.797 |
| | Innovated form of live streaming | 0.483 |
| | Being able to access to previously unknown knowledge | 0.644 |
| Aesthetics Alpha = .39 | The UI and visual design of the app | 0.566 |
| | Streamer's appearance | 0.813 |
| Communication with Others Alpha = .77 | Sending and reading comments in the stream | 0.754 |
| | Communicate with peer viewers | 0.794 |
| | Keep in touch with the streamer | 0.576 |
| | Joining the streamer's fan group | 0.589 |
| Emotional Reactions Alpha = .69 | Being able to feel streamer's emotion | 0.463 |
| | Improvisation and uncertainty of the stream | 0.635 |
| | Sending gifts to the streamer | 0.766 |
| | Leading in the leaderboard of gift-senders | 0.736 |

**Table 3. Factor Analysis with 4 factors derived**

**DISCUSSION AND IMPLICATIONS FOR DESIGN**
We now reflect on our findings of how live streaming apps were used in China and compare our results with other research on live streaming. We provide insights to inform future research and design in the HCI community.

Our results echoed those found prior in that viewers liked the authentic, unedited nature of live streaming [33] and the community fostered from the streams [8]. The content is also less important than the way streamer performs it, as found in [31] that "content structure is king". However, live streaming practices in China differ in several ways.

One big difference is that streams about politics [7] or civic content [4] are rare in China. Although some streamers sometimes banter about current affairs, such content constitutes a very small portion of their streams, and they do so more for gaining attention than for informing viewers of civic content. We conjecture that censorship is one reason, as found in [21] that people in China seldom post about politics to "avoid placing themselves in potentially difficult situations". All live streaming platforms in China require users to verify their identity before going live or commenting, which may have a dampening effect. Another reason may be that young Chinese people tended not to be interested in democratic political systems [19,41]. Interviewees also noted that WeChat or Weibo are used more for following news, while live streaming is more for pure entertainment or learning.

The interviewees perceived little negative content, though reported trolls and inappropriate content being streamed on public channels. They noted that trolls are often blocked by the streamers or their moderators, and inappropriate content is often cut off by the official platform regulators. They also viewed regulations and censorship as positive forces to reduce content that may be undesirable or negative.

We found motivations for using live streaming aligned with the "hot and noisy" principle ("*renao*") that Chinese people use social media for "achieving a harmonious ambience of social life" [36]. People use live streaming for more pleasure, more chances for socializing with others, and a feeling of being trusted and accompanied by others. The interviewees also mentioned having added streamers or peer viewers as their contacts on QQ or WeChat who were strangers before, showing the willingness to befriend and communicate with strangers online [36]. Live streaming "*provides a perfect place for befriending strangers in a socially acceptable way*" (P1).

We also noticed different attitudes towards watching friends' streams. Previous work shows that when viewing live events, people interact more on friends' live streams than strangers' [7,18]. However, when asked about watching friends' streams, some participants said they preferred to watch more popular streamers who were strangers, especially if their friends just showed their daily life, although they would like to support and trust their friends. The reasons they mentioned were their needs to keep a distance and respect their privacy, familiarity and potential bore of the streams, having other channels for better interactions, and awkwardness that might incur.

Sending virtual gifts and forming fan groups may be rooted in some cultural differences. Previous research has shown that Asian users prefer multi-party chat, audio-video chat, and IM emoticons [14], which may be why they use private chatroom services that support multi-modal messaging for community building rather than Facebook or Twitter-like public social networks. Chinese people think highly of *guanxi* [40] in social interactions, which may be why they reward streamers with virtual gifts. They may perceive the rewarding process not only as a consumer behavior, which can be an impulse purchase, but also a social interaction for circulating *guanxi* and keeping 'face' [40]. Virtual gifts also act as channels supporting social dialogues [16], 'currency' of social capital [40], and souvenirs [34].

Our findings echo previous research about live streaming as a virtual *third place* [8], or informal public spaces where people engage in socializing with others and form communities [25]. We also show that fan groups are another *third place*. Live streams are open to all and public, with synchronized communication, and to some extent, ephemeral messages (i.e., once the stream is over, the chat cannot go on in context and hard to resume), while fan groups are closed and private, with both synchronized and asynchronized communication, permanent messages, and multi-modal channels, which complement the nature of live streaming. Viewers keep engaging in these additional third places and seem to build more trust from communication with each other, e.g., they begin to disclose personal information or even meet up in person offline.

Based on our results, we now provide insights about how to design for better social live streaming experiences.

### Supporting Deeper and Richer Interactions
The practice of leveraging fan groups for casual communications within the community indicates that viewers desire deeper and richer interactions with other viewers and streamers. Besides multi-modal messages, some participants even desire to convey the feeling of applauding or remotely hugging the streamer, or being able to feel, taste or smell objects in the stream. Solutions to how to make such interactions less intrusive and interruptive, and how to control trolls in the form of new modality, will be challenging to create. For example, future work can explore how to design customizable multi-modal gifts to support richer social interactions in live streaming.

### Live Streaming within Non-Entertainment Contexts
Our results showed that live streaming has penetrated other services, such as e-commerce and online education. As the current design of live streaming applications is to support entertainment it remains to be seen how to best redesign live streaming apps to support the unique challenges of these diverse experiences while still maintaining a sense of community, reward, and authenticity. For example, for educational streams, it suffers from low efficiency and requires intense preparation for streamers. Future work can explore how to make it more efficient, and how to design tools to engage online learners and the crowd and support collaborative preparation of live informational content.

### Leveraging Live Streaming for Social Good
We found that some streamers have already begun to use live streaming to showcase traditional cultural art forms and artifacts. We envision this as a chance to preserve endangered cultures by exposing them to a broader audience to raise cultural awareness. Knowledge and experience sharing streams are also beginning to attract older adults to watch and engage in live streaming. We should thus design better social interactions for these older adults to feel less excluded from live streaming, and design community-based mechanisms that would engage them.

### CONCLUSION
We empirically examined user's live streaming practices in China to better understand such social-technological phenomenon, which has a large and ever-growing market, and different content types, compared to those in North America. By understanding the motivations, practices, social interactions that occur beyond the streams, and the important factors of engagement, we identified that the unique challenges and viewing behaviors encountered with Chinese live streams can be used as blueprint for the future of North American and European live streaming platforms and services. Our work also outlined the importance of expanding thinking about the current demographics of users, live streaming within educational and finance applications, and the needs to support deeper and richer interactions between viewers and streamers.